\documentclass[aps,prl,reprint,10pt]{revtex4-1}

\usepackage{graphicx} 
\usepackage{comment}
\usepackage{mathtools}
\pdfoutput=1
\begin{document}
\title{Strong CP solution with soft PQ breaking}

\author{Ravi Kuchimanchi}
\email{ravikuchimanchi1@gmail.com}
\begin{abstract}

A recent work combined the popular left-right parity (LR)  and Peccei-Quinn (PQ) symmetries to explain the alignment in quark masses. 
Since axions may not exist, we break PQ softly and discover a new solution to the strong CP problem.  Remarkably, since the soft PQ breaking terms respect parity, the strong CP phase  is absent at  tree-level, while  $\bar{\theta} \gtrsim 7 \times 10^{-11}$ is generated in two loops. Thus a neutron electric dipole moment close to the  current experimental bound is predicted, making the theory falsifiable, even if the scale of new physics is well beyond the reach of colliders. In LR models leptonic CP phases (such as Dirac phase $\delta_{CP}$) can generate the strong CP  phase in one loop, which is a challenge for their existence. PQ symmetry sets one of the leptonic Yukawa matrices to zero at the tree level and automatically suppresses this contribution, and makes a way for leptonic CP violation to be present.

\end{abstract}

\maketitle

\noindent \emph{Introduction -} 
 While Peccei-Quinn (PQ) symmetry~\cite{PhysRevLett.38.1440} is the most celebrated and cerebrated solution to the strong CP problem, parity (P)
 restores the balance between left and right, predicts the existence of right-handed neutrinos, and historically anticipated neutrino masses and mixings~\cite{PhysRevD.10.275,PhysRevD.11.566,Senjanovic:1975rk}.  A recent work~\cite{Dev:2018pjn}  combines these  popular symmetries in an interesting way  to explain the alignment in the hierarchy of quark masses.  

Since $U(1)_{PQ}$ has an anomaly, there must be pseudo-goldstone bosons or axions with a tiny mass. 
 However so far  no experimental evidence for axions has been found.    
  Moreover, since PQ is anyway not an exact symmetry, adding explicitly terms that would break it softly is a natural idea to consider, which would give a large mass to the would-be axion. However this would generally spoil the strong CP solution, by generating an $O(1)$ contribution to the strong CP phase $\bar{\theta}$.

But remarkably when PQ symmetry is combined with parity, we find that $\bar{\theta}$ is not generated at the tree-level, since the minimum for the would-be axion field, on soft PQ breaking is parity symmetric. The soft breaking terms also do not spoil the alignment of quark masses discussed in~\cite{Dev:2018pjn}, thereby giving a larger canvass of theory space for it.

PQ symmetry sets one of the quark Yukawa coupling matrices ($\tilde{h}_q$), needed for CKM mixing, to zero. In reference~\cite{Dev:2018pjn}, in order to radiatively generate ($\tilde{h}_q$)  below the spontaneous PQ breaking scale, inspired by the work in~\cite{Balakrishna:1987qd,Balakrishna:1988ks,Babu:1989tv},  color triplet scalars   that couple to the quarks were used. What we find is that these same couplings also radiatively generate $\bar{\theta}$ in two loops, if PQ symmetry is broken softly.  

Since the same couplings that generate the CKM  mixing also give rise to $\bar{\theta}$ it turns out to be in a range that can be experimentally observed.  We find $\bar{\theta} \gtrsim 7 \times 10^{-11}$ is generated which implies that the neutron electric dipole moment may be detected earlier rather than later. Since experiments with sensitivity to probe $\bar{\theta} \sim 10^{-12}$ are in advanced stages of preparation,  the idea of softly broken PQ symmetric model protected by parity is falsifiable, independent of the scale of new physics which can be well beyond collider reach. Note that the words `strong CP solution', imply that there are no unnatural cancellations of $\bar{\theta}$, which we assume.  Quadratic divergences are dealt with the usual way, and contain no imaginary or CP violating component.  

Planck scale effects through non-renormalizable operators can generate a worrisome $\bar{\theta}$. Since Majorana mass of the right-handed neutrino ($\nu_R$) breaks lepton number, in LR models the seesaw scale is generally identified with $B-L$ gauge symmetry breaking scale $v_R$. Due to PQ symmetry, neutrino's Dirac mass is generated from the $\tau^-$ Yukawa coupling, making the seesaw scale $\sim 10^7 GeV$. There are no gauge singlet scalars, and  $\bar{\theta}_{Planck} \sim (v_R/M_{Pl})^2 \sim 10^{-22}$. Thus an understanding of the smallness of Planck scale corrections emerges. 

Another key  reason to take this idea seriously is that left-right symmetric models (LR) also have a leptonic CP problem~\cite{Kuchimanchi:2014ota} -- CP violating phases in product of Dirac and Majorana type Yukawa couplings can generate the strong CP phase in just one loop.  So a worthy axionless solution to the strong CP problem must also address this problem. Just as the PQ symmetry sets a quark Yukawa coupling matrix to zero,  it also sets a Dirac type leptonic Yukawa coupling matrix to zero, and the contribution from the product of leptonic Yukawa matrices vanishes automatically at the tree-level. 

There is already a solution to the strong CP problem in the LR model where in addition to P, the Lagrangian is also made CP symmetric~\cite{Kuchimanchi:2010xs}. This solution can set all the leptonic CP phases to zero~\cite{Kuchimanchi:2012te}, and thereby also address the leptonic CP problem.  This would imply that leptonic $\delta_{CP}$ phase being measured by neutrino experiments will turn out to be consistent with $0$ or $\pi$.

Moreover, if there are no axions, in most of the parameter space of the minimal LR model itself, $\delta_{CP}$ would need to vanish (mod $\pi$) as too high a strong CP phase is generated in one loop from the leptonic Yukawa matrices~\cite{Kuchimanchi:2014ota}.   

The axionless solution using PQ and P symmetries is important as it would explain why a leptonic Yukawa matrix vanishes at the tree-level, thereby providing a \emph{raison d'etre} for picking a  special region of parameter space where leptonic $\delta_{CP}$ can be present.  
That it is falsifiable by neutron EDM experiments is extremely fortuitous.



\vspace{0.3cm}
\noindent \emph{Softly broken PQ symmetry - }
\vspace{0.3cm}

 We consider the Left-Right PQ symmetric model $(LRPQ)$ based on $ SU(3)_c \times SU(2)_L \times SU(2)_R \times U(1)_{B-L} \times PQ \times P$ that was studied in reference~\cite{Dev:2018pjn}.  Analogous to $SU(2)_L$ doublets $Q_{iL} (3, 2, 1, 1/3;1) $ and $L_{iL} (1, 2, 1, -1;-2)$, the right handed quarks $Q_{iR}(3, 1, 2, 1/3;-1)$ and leptons $L_{iR} (1, 1, 2, -1;0)$ are doublets of $SU(2)_R$,   where i = 1, 2, 3 is the generation index, that we do not henceforth write explicitly.  The first four numbers inside each of the  round brackets are the respective gauge charges, while the last number indicates the global $U(1)_{PQ}$ charge.  

Due to the PQ symmetry, one of the two usual Yukawa coupling matrices that give masses to the up and down sector quarks is absent. Therefore there will be no CKM mixing unless we allow for an additional coupling. We do this by adding a set of scalar color-triplets $\omega_L (3, 1, 1, -2/3;-2)$ and $\omega_R (3, 1, 1, -2/3;2)$ that can interact with the quarks~\cite{Dev:2018pjn}. The remaining scalar fields are the usual ones of the minimal left right symmetric model, consisting of  $\Delta_L (1,3,1,2;4)$ and $\Delta_R (1,1,3,2;0)$, and bi-doublet $\phi (1,2,2,0;2)$. We do not require gauge singlet scalar fields of reference~\cite{Dev:2018pjn} as we will break the PQ symmetry softly rather than spontaneously.

Under parity (P),  the space-time coordinates $(x,t) \rightarrow (-x,t), \ \phi \rightarrow \phi^\dagger$ and subscripts $L \leftrightarrow R$ for all other fields.   We can represent the $SU(2)_L$ or $SU(2)_R$ doublet fermions as column vectors of the respective gauge group (such as $Q_{3L} = (t~~b)^T_L$ or $L_{3R}= (\nu_\tau~~\tau)^T_R$ ) and scalars:  

\begin{equation}
\begin{array}{ccc}
\phi = \left(\begin{array}{cc}
\phi^o_1 & \phi^+_2 \\
\phi^-_1 & \phi^o_2
\end{array}
\right), 
 &
\Delta_{L,R} = \left(\begin{array}{cc}
\delta^+_{L,R} / \sqrt{2} & \delta^{++}_{L,R} \\
\delta^o_{L,R} & - \delta^+_{L,R} /\sqrt{2}
\end{array}
\right), 
\end{array}
\label{eq:fields}
\end{equation}
 There are two standard model Higgs doublets in the bidoublet $\phi$,  indicated by the subscripts $1$ and $2$. The second  doublet is naturally heavy with mass at the $SU(2)_R \times U(1)_{B-L}$ breaking scale $\left<\delta^o_R\right> =v_R >> \left<\delta^o_L\right>$, where parity also breaks spontaneously. 
 $v_R$ can be anywhere between a few $TeV$ and the Planck scale. 


To simplify the discussion, we will break LRPQ to the Standard model in two steps, $LR \times PQ \xrightarrow{\mu_{PQ}} LR \xrightarrow{v_R} SM$ (the reverse order of breaking is considered later).

First we break the PQ symmetry at a high scale $\sim \mu_{PQ}^2 >> v_R^2$ by adding the following soft PQ breaking terms to the Higgs potential 
\begin{equation}
\label{eq:softpq}
\mu^2_{PQ} \omega_L^\dagger \omega_R + \mu_2^2 \tilde{\phi}^\dagger \phi + hc
\end{equation}
where $\tilde{\phi} = \tau_2 \phi^\star \tau_2$.  Note that $\mu_2^2$ and $\mu_{PQ}^2$ are real due to P and we take $\mu_{PQ}^2$ to be positive without loss of generality. At the higher PQ breaking scale the only other relevant mass terms are those that give masses to $\omega_L$ and $\omega_R$ namely,
	\begin{equation}
	\label{eq:omegamass}
	m^2 (\omega_L^\dagger \omega_L + \omega_R^\dagger \omega_R)
	\end{equation}

We can now go to the mass basis by defining 
\begin{equation}
\label{eq:omega}
\omega_\pm = (\omega_L \pm \omega_R)/\sqrt{2}
\end{equation}
so that the mass terms in eqns~(\ref{eq:softpq}) and~(\ref{eq:omegamass}) involving the scalar color triplets can be rewritten as:
\begin{equation}
m_+{^2} \omega_+^{\dagger}\omega_+ + {m_-}^2 \omega_-{^\dagger}\omega_-
\end{equation}
where ${m_\pm}^2 = m^2 \pm \mu_{PQ}^2$.  Note that under $P$, $\omega_\pm \rightarrow \pm \omega_\pm$ (since $\omega_L \leftrightarrow \omega_R$) and these terms do not break parity.

Since we took $\mu^2_{PQ} \geq 0$, the color-triplet scalar  $\omega_+$ is heavier than $\omega_-$ and decouples first.  Also, ${m_\pm}^2 > 0$ so that QCD remains unbroken, and $m^2 \sim \mu_{PQ}^2$ so that $Ln(m_+/m_-) \approx 1$. Note values of $Ln(m_+/m_-)$ larger than $1.5$ wound need $\mu_{PQ}$ to be tuned very close to $m$.  

Below ${m_+}^2$, since $\omega_+$ decouples, we have just the minimal LR model with an additional color triplet $\omega_-$.  Therefore between the scales $m_+$ and $m_-$ the Yukawa couplings involving the quarks are given by those of the minimal LR model with an additional term due to $\omega_-$ as below,

\begin{eqnarray}
\label{eq:yuk}
h_q \bar{Q}_{L} \phi Q_R + \tilde{h}_q  \bar{Q}_{L} \tilde{\phi} Q_R  + {\frac{g}{\sqrt 2}} (Q_L^T \tau_2 \omega_- C^{-1} Q_L + h.c.) \nonumber \\+ L\leftrightarrow R, \phi \rightarrow \phi^\dagger, \omega_- \rightarrow -\omega_- \ \ 
\end{eqnarray}
where $h_q$ and $\tilde{h}_q$ are $3\times 3$ matrices in generation space and Hermitian due to parity, while $g$ is a complex symmetric $3\times 3$ matrix. Note that the Yukawa matrix $g$ is the same as the one in the PQ symmetric term $g Q_L^T \tau_2 \omega_L C^{-1} Q_L + g Q_R^T \tau_2 \omega_R C^{-1} Q_R$, considered in  reference~\cite{Dev:2018pjn}, with $\omega_L$ and $\omega_R$ written in terms of $\omega_\pm$  using eqn.~(\ref{eq:omega}), and dropping the $\omega_+$ term as it decouples.   

Importantly at the scale $m_+$, $\tilde{h}_q$ vanishes as it is not permitted by PQ symmetry above this scale. However below this scale,  $\tilde{h}_q$  gets generated from $h_q$ and $g$ in one loop due to RGE running between $m_+$ and $m_-$, as discussed in reference~\cite{Dev:2018pjn}.

\begin{equation}
\label{eq:RGEh}
{{d\tilde{h}_q} \over {d{ln \mu}}} \sim {\frac{\left(g^\dagger h_q g\right)}{16\pi^2}}
\end{equation}
Working in a basis where $h_q$ is diagonal, the CKM mixing depends on off-diagonal terms in $\tilde{h}_q$.  Note that in order to generate sufficient CKM mixing or $\tilde{h}_q$ through RGE running, the Yukawa matrix $g$ must have sufficiently large entries. 


We will now discuss the strong CP problem in this model. Parity sets $\theta_{QCD}$ to zero and ensures that $h_q$ and $\tilde{h}_q$  are Hermitian.  As noted in reference~\cite{Dev:2018pjn} all parameters of the Higgs potential turn out to be real at and above the PQ symmetry breaking scale.  However while it was noted that parameters were real, the connection that this could lead to the resolution of strong CP problem without an axion was missed in reference~\cite{Dev:2018pjn}.  

We also note that the soft PQ breaking terms given in eqn~(\ref{eq:softpq}) are both real.   Thus though CP is broken by complex phases in $g$, all terms of the Higgs potential are real at and above the scale $m_+$.  The Higgs VEVs obtained by minimizing the Higgs potential are all automatically  real or CP conserving at this scale.
     
As can be seen from eqns.~(\ref{eq:fields}) and (\ref{eq:yuk}), the strong CP phase $\bar{\theta} = Arg Det (M_u M_d)$ vanishes at tree-level, since the up and down quark mass matrices $M_u$ and $M_d$ turn out to be Hermitian as they are determined by Hermitian $h_q$ (and $\tilde{h}_q$) multiplying the bidoublet Higgs VEVs $\left<\phi_1^o\right>=\kappa$ and $\left<\phi_2^o\right>=\kappa'$ that are real at tree-level. $|\kappa|^2 + |\kappa'|^2 = (174 GeV)^2$ is the weak scale and $tan \beta = |\kappa_1/\kappa_2|$.

It is important to note that PQ symmetry plays a crucial role in the above reasoning.   Without PQ symmetry there can be  the CP violating Higgs potential term  
\begin{equation}
\label{eq:alpha2}
\alpha_2 Tr(\tilde \phi^\dagger \phi \Delta_R^\dagger \Delta_R) + (R \rightarrow L, \phi \rightarrow \phi^\dagger) + h.c.,
\end{equation} 
where $\alpha_2$ is complex and can generate $\bar{\theta}$.  However this term violates $U(1)_{PQ}$ and $\alpha_2 = 0$ at and above the scale $m_+$. 

\begin{figure}[t]
\begin{center}
\includegraphics[height=2cm]{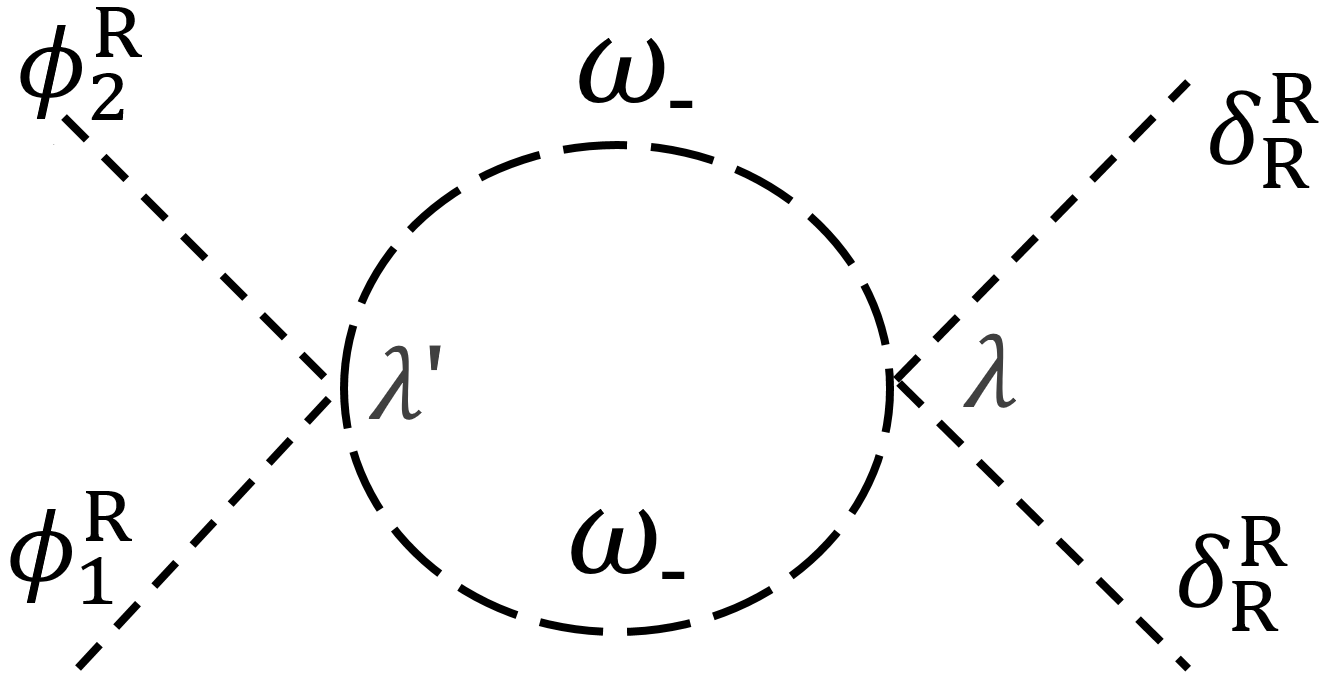}
\end{center}
\caption{One loop contribution that generates $\alpha_2$. Had PQ not been broken the masses of $\omega_+$ and $\omega_-$ would not be split. Then  a similar diagram with $\omega_+$ in the loop that couples with the bidoublets with $-\lambda'$ instead of $\lambda'$ would exactly  cancel the above contribution. However since PQ is broken softly, $\omega_+$ is heavier with mass $m_+$ and decouples. Since $\lambda, \lambda'$ are real this does not generate $\bar{\theta}$. The fields shown are defined in caption of figure~\ref{fig:twoloop}.}      
\label{fig:oneloop}
\end{figure}

Like $\tilde{h}_q$, it gets generated below the PQ breaking scale $m_+$ due to RGE running. However the one loop contribution to  $\alpha_2$  from figure~\ref{fig:oneloop} is real and CP conserving and therefore does not generate $\bar{\theta}$. Note that we have used the Higgs potential terms
\begin{eqnarray}
\label{eq:higgs}
Tr( \lambda' \tilde{\phi}^\dagger \phi + \lambda  \Delta_R^\dagger \Delta_R)  \omega_-^\dagger \omega_-   + R \rightarrow L, \phi \rightarrow \phi^\dagger
\end{eqnarray}
where the first term is from the PQ symmetric term  $ Tr(\tilde{\phi}^\dagger \phi) \omega_R^\dagger \omega_L$ written in terms of $\omega_\pm$ with $\omega_+$ having decoupled. Since this term was missed in reference~\cite{Dev:2018pjn}, we note that $\lambda'$ is real due to parity. The second term of eqn.~(\ref{eq:higgs}) is from PQ symmetric terms $\omega_R^\dagger \omega_R Tr (\Delta_R^\dagger \Delta_R)$ and $\omega_L^\dagger \omega_L Tr (\Delta_R^\dagger \Delta_R)$ (and $L \leftrightarrow R$). 

\begin{figure}[t]
\begin{center}
\includegraphics[height=2.8cm]{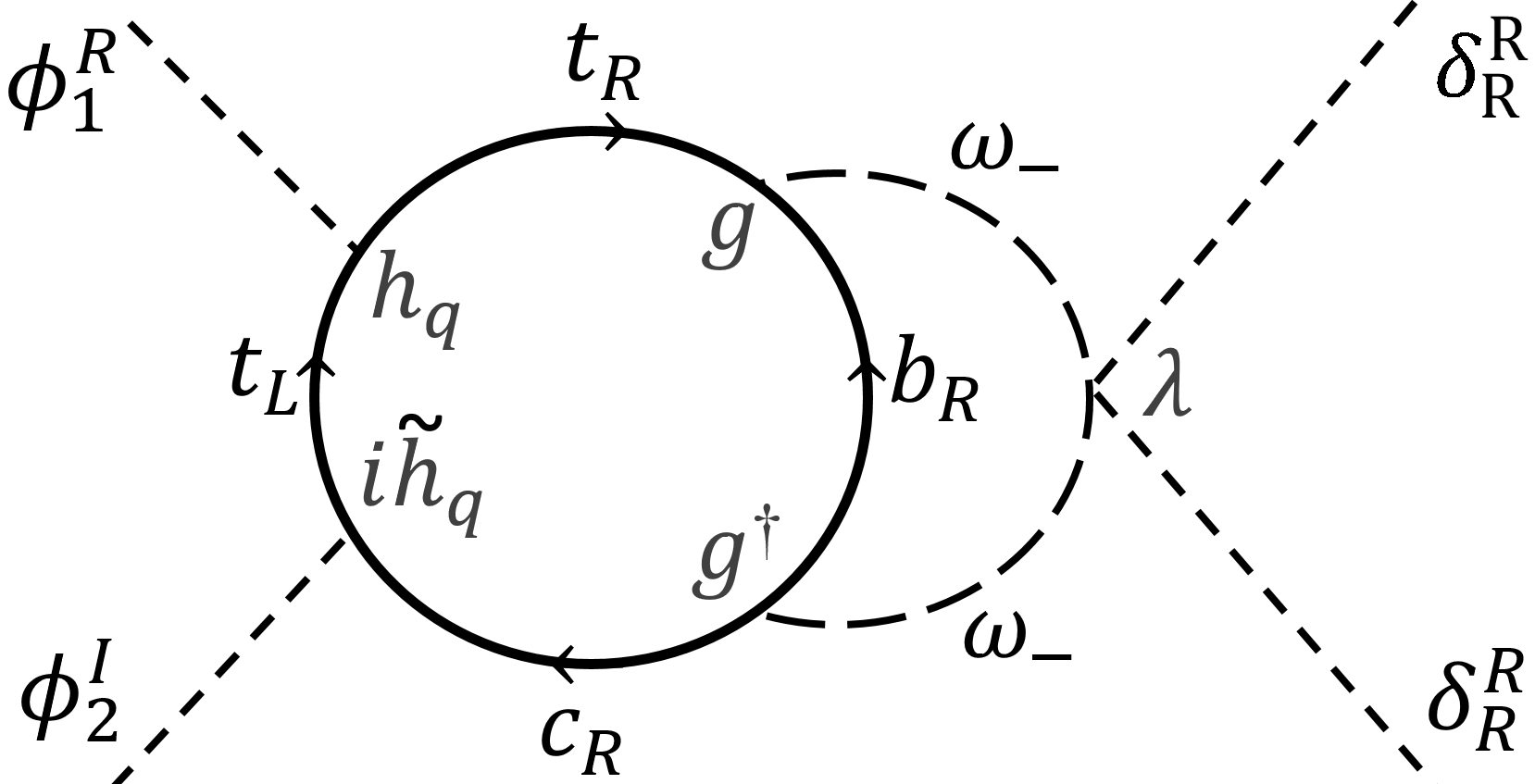}
\end{center}
\caption{Two loop contribution to renormalization group running of $\alpha_{2I}$, the imaginary part of $\alpha_2$ of eqn.~(\ref{eq:alpha2}). Shown in the figure are the neutral components of scalar fields in eqn.~(\ref{eq:fields}), $\phi_{1,2}^o = \phi_{1,2}^R + i \phi_{1,2}^I$ and $\delta^o_R = \delta_R^R + i \delta_R^I$ with superscripts denoting the real and imaginary parts. Here the PQ breaking term is $\tilde{h}_q$  given by eqn.~(\ref{eq:RGEh}). The quark loop generates $i Tr(h_q \tilde{h}_q g^\dagger g)$. There is also a diagram with $\phi_1^R$ and $\phi_2^I$ crossed that generates $-i Tr(\tilde{h}_q h g^\dagger g)$ leading to equation~(\ref{eq:RGEalpha}).     }   
\label{fig:twoloop}
\end{figure}

From figure~\ref{fig:twoloop} we find a non-vanishing two-loop contribution to $\alpha_{2I}$ (the imaginary part of $\alpha_2 = \alpha_{2R} + i \alpha_{2I} $) leading to the RGE
\begin{equation}
\label{eq:RGEalpha}
{{d\alpha_{2I}} \over {d{ln \mu}}} \sim {\frac{6 \lambda i}{\left( 16 \pi^2 \right)^2}}   \left[   Tr\left(  \left[ h_q, \tilde{h}_q \right] g^\dagger g  \right) \right]
\end{equation}
where we have used the Yukawa couplings from eqn.~(\ref{eq:yuk}) and a Higgs quartic coupling from eqn~(\ref{eq:higgs}). The factor of $6$ is obtained from top and bottom quark and $\omega_-$ colors  in the loops, and from a factor of $1/2$ from  $g/\sqrt{2}$. Note that $i[h_q,\tilde{h}_q]$ is a Hermitian matrix.

We now solve the RGEs~(\ref{eq:RGEh}) and~(\ref{eq:RGEalpha}) together by integrating from $m_+$ to $m_-$ with the boundary condition that $\tilde{h}_q=0$ and $\alpha_2=0$ at renormalization scale $\mu = m_+$.  We work in a basis where $h_q$ is diagonal so that its  entries are just the top, charm and up Yukawa couplings. We allow for slightly smaller quark masses  since soft PQ breaking terms and the mechanism to generate CKM mixing could be at a high scale such as the seesaw, GUT or Planck scale. We can have for example at the GUT scale~\cite{Bora:2012tx,Das:2000uk}, $m_t \sim 70 GeV$ and $m_b \sim 0.9 GeV$ corresponding to Yukawas $h_t \sim 0.4$ and $h_b \sim 0.006$.   Values of $g$ along with $m_t/m_b = tan \beta$ are chosen to generate sufficient CKM mixing and the bottom sector quarks' masses.   As already discussed,  we take $Ln (m_+/m_-) \approx 1$. 

We use R program to integrate the two RGEs and estimate that $\alpha_{2I} \gtrsim 10^{-9}$  is generated. A quick back of the envelope estimate of eqn~(\ref{eq:RGEalpha}) gives a similar figure $\alpha_{2I} \sim 6 \lambda/(16\pi^2)^2 (h_t h_b V_{cb} g_{23} g_{33})$ with $h_t \sim 0.4, h_b = 0.006,$ CKM mixing $V_{cb} \sim 4 \times 10^{-2}$ and  $g_{23} g_{33} \sim 0.1$. 



Using eqns~(\ref{eq:alpha2}) and (\ref{eq:fields}) we see that once $\delta^o_R$ picks up a large VEV $v_R$ and breaks $P$, $\alpha_{2I}$ generates an imaginary part in the VEV $\kappa'$ of  the second Higgs doublet in bi-doublet $\phi$.  This makes the up and down quark mass matrices non-Hermitian and generates a strong CP phase given by
\begin{equation}
\label{eq:thetaalpha}
  \bar{\theta} = (\alpha_{2I}/\alpha_3) (m_t/m_b) \approx 7 \times 10^{-8} \times \lambda 
\end{equation}
where the first equation is from eqn. (3) of reference~\cite{Kuchimanchi:2014ota}, and $\alpha_3$ is a Higgs quartic coupling that gives the mass $\alpha_3 v_R^2$ to the second standard model doublet in the bidoublet $\phi$. We have substituted $\alpha_{2I} \gtrsim 10^{-9}$ and $m_t/m_b \sim 70$, and took $\alpha_3 \sim 1$. 


 The Higgs quartic coupling in eqn.~(\ref{eq:higgs}), $\lambda \sim g_{B-L}^4 ln(\mu_{PQ}/\Lambda)/(16 \pi^2)$  is generated in one loop from $U(1)_{B-L}$ gauge bosons, where $\Lambda$ is a cut-off such as the Planck scale.  Allowing for the possibility that soft PQ breaking terms could be near the cut-off  scale we obtain  $\lambda \gtrsim 10^{-3}$.     

Substituting in eqn.~(\ref{eq:thetaalpha}) leads us to the prediction,
\begin{equation}
\bar{\theta}\gtrsim 7 \times 10^{-11}
\end{equation}

This is exciting because it means that the neutron EDM experiments should be able to detect a positive signal soon, or rule out the idea of softly broken PQ symmetry protected by parity as discussed in this work.   

We also make a note of the current bounds.  Estimates for neutron EDM are $d_n = (-2.7 \pm 1.2) \times 10^{-16} \bar{\theta}~ecm$ from chiral effective field theory~\cite{Yamanaka:2017mef}, while a lattice QCD analysis gives $d_n = -0.90(15) \times 10^{-16} \bar{\theta}~ecm$~\cite{Shindler:2015aqa}. The current experimental bound  $d_n \leq 3.6 \times 10^{-26} ecm$ at $95\%$ confidence level~\cite{Afach:2015sja} translates to $\bar{\theta} \leq 2.4 \times 10^{-10}$ based on the first estimate, while the second implies $\bar{\theta} \leq 4 \times 10^{-10}$.  

We now consider the case where LRPQ  breaks in the reverse order, that is $ LR \times PQ \ \ {\stackrel{v_R}{\longrightarrow}} \ \ SM \times PQ \ \ {\stackrel{\mu_{PQ}}{\longrightarrow}} \ \ SM $.   

In this case below $v_R$, there is no parity to protect the Hermiticity of Yukawa matrix $\tilde{h}_q$ while it is being generated from $h_q$ and $g$. 

\begin{figure}
\begin{center}
\includegraphics[height=2.5cm]{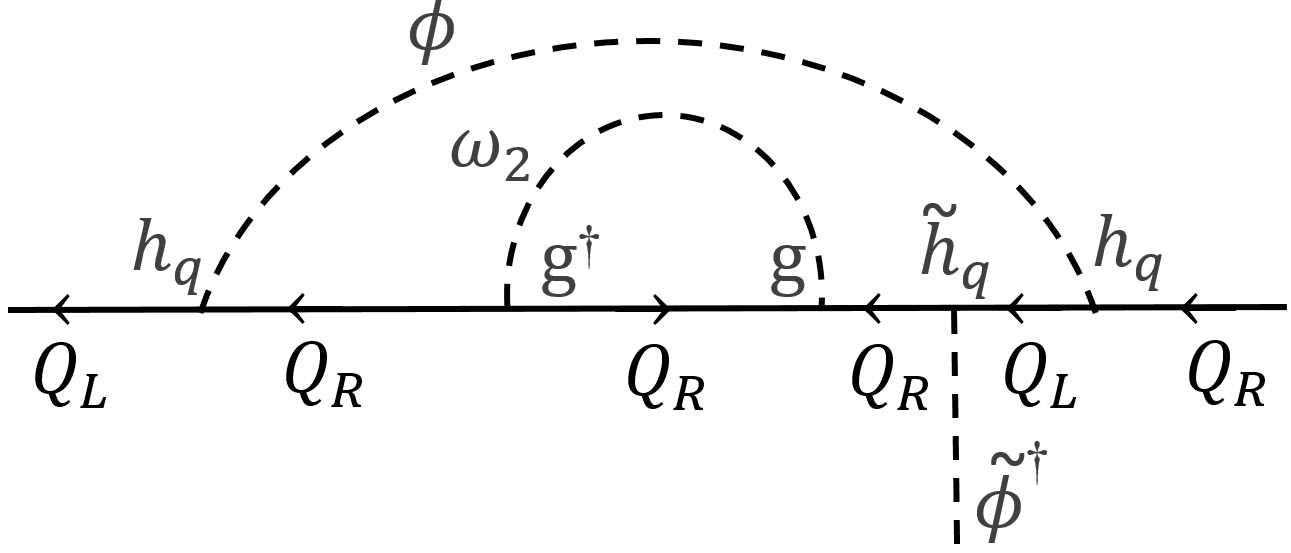}
\end{center}
\caption{Two loop contribution to non-Hermiticity of $\tilde{h}_q$ below parity breaking scale $v_R$. A factor of $cos^2 \alpha$ multiplies the $g^\dagger g$, where $\alpha$ is the mixing between $\omega_L$ and $\omega_R$. There is a similar diagram with the external scalar leg placed on the left side, but it's matrix structure is the Hermitian conjugate with a  multiplying factor $\sin^2 \alpha$.  $\lambda (\omega_R^\dagger \omega_R) v_R^2$ that gets generated  on parity breaking causes the mixing to be non-maximal with $(cos^2\alpha - sin^2\alpha) \sim \lambda (v_R/\mu_{PQ})^2$ and generates $\bar{\theta}$. $\omega_2$ is the lighter mass eigenstate of $\omega_L$ and $\omega_R$. }   
\label{fig:twoloop-case2}
\end{figure} 

The mathematical reason why $\bar{\theta}$ was generated from figure~\ref{fig:twoloop} is that with Hermitian $h_q$ and complex symmetric $g$, the product $h_q (g^\dagger h_q g) g^\dagger g$  is not Hermitian and has a trace that is imaginary. Note that  $\tilde{h}_q$ in figure~\ref{fig:twoloop} is like $g^\dagger h_q g$ due to eqn.~(\ref{eq:RGEh}).  As can be seen from figure~\ref{fig:twoloop-case2} and its caption, similar products of Yukawa matrices also occur in the two-loop contribution to $\tilde{h}_q$ making it non-Hermitian below the parity breaking scale for the case where $v_R^2 > \mu_{PQ}^2$.  Thus a similar $\bar{\theta}$ as in the previous case is generated.   

\vspace{0.3cm}
\noindent \emph{Leptonic CP, Planck scale and conclusions - }
\vspace{0.3cm}

In the minimal LR model $Im Tr(h^\ell \tilde{h}^\ell f^\dagger f) \leq 3 \times 10^{-11}$ as this trace generates $\alpha_{2I}$ in one loop RGE and contributes to $\bar{\theta}$~\cite{Kuchimanchi:2014ota}. Here $h^\ell, \tilde{h}^\ell$ are Hermitian Dirac Yukawa matrices for the leptons, and $f_{ij}$ is the Majorana Yukawa matrix.  This implies that in most of the regions of parameter space we may naively consider, the leptonic CP phases must vanish.  This is also backed by an ultraviolet completion of the LR model with an  axionless solution that imposes both P and CP~\cite{Kuchimanchi:2010xs}.   On CP breaking only the CKM phase is generated in the minimal version, and not leptonic CP phases~\cite{Kuchimanchi:2010xs,Kuchimanchi:2012te}.  This solution is testable by its prediction that leptonic $\delta_{CP} = 0$ mod $\pi$.

Our new solution provides a different way of addressing the leptonic CP issues.   The PQ charges have been assigned to leptons and quarks such that $h^{\ell} = \tilde{h}_q = 0$.  Thus the above trace vanishes at the tree level, while leptonic CP phases can be present.  Moreover since the largest Yukawa coupling of leptons is of $\tau^- \sim 10^{-2}$, the off-diagonal terms of $h^\ell$ radiative generated  on PQ breaking are negligibly small. 
The new solution  selects the special region near $h^\ell = 0$ where  $\delta_{CP}$ can be present. It is testable by its strong CP phase prediction.   

Thus the left-right symmetric model without an axion will be put to non-trivial testing in the next few years through neutrino and neutron EDM  experiments. The presence of $\bar{\theta}$ in the next level of sensitivity or absence of $\delta_{CP}$ violation would hint at parity being restored in the basic laws of nature, even if it is at a scale we may never be able to scale in any other way.

Since $h^{\ell} = 0$ at tree-level, the Dirac mass to the neutrino is generated from $\tilde{h}^\ell$ and is $\sim m_\tau tan\beta \approx m_\tau (m_b/m_t)$.  Together with observed light neutrino masses this implies that the seesaw scale $f_{33} v_R \sim 10^7 GeV$.   Since there are no gauge singlets, the $\bar{\theta}$ from Planck scale contribution to quark masses is suppressed by $\sim v_R^2/M_{Pl}^2 \sim 10^{-22} /f_{33}^2$, with $M_{Pl} \sim 10^{18} GeV$. If lepton number breaks at the $B-L$ gauge symmetry breaking scale, as is usually assumed on physical grounds,  $f_{33} \sim 1$ and these corrections are negligible. The suppression factor is $\leq 10^{-10}$ all the way up to $v_R \sim 10^{13} GeV$ with $f_{33} \sim 10^{-6}$. It's interesting that these corrections do not depend on $\omega_\pm$ masses and $\mu^2_{PQ}$, which could be near the Planck scale.

In case non-trivial $\delta_{CP}$ is discovered, while  $\bar{\theta}$ at levels predicted in this work is not found, then we can still solve the strong CP problem using~\cite{Kuchimanchi:2010xs}, and adding a generation of vectorlike leptons~\cite{Kuchimanchi:2012te} below the P breaking scale $v_R$.  Their couplings can generate a CP phase in the leptonic sector, after LR symmetry breaks, thereby evading the leptonic CP problem. Vector like quarks that generate the CKM phase can be above or below $v_R$. In this case these vectorlike fermions could be within reach of colliders, though we would not have the minimal LR model at any scale.

We have stumbled upon a remarkable solution to the strong CP problem, that we obtain by simply adding soft PQ breaking terms to the  left-right PQ symmetric model. The solution predicts $\bar{\theta} \gtrsim 7 \times 10^{-11}$, independent of the scale of new physics. The idea of soft PQ breaking protected by parity  is falsifiable by neutron EDM experiments.

\bibliography{pqlr_bibtex}

\end{document}